# Model of deep non-volcanic tremor part I: ambient and triggered tremor

Naum I. Gershenzon, Gust Bambakidis

Physics Department & Department of Earth and Environmental Sciences, Wright State University, 3640 Colonel Glenn Highway, Dayton, OH 45435

**Abstract**

There is evidence of triggering of tremor by seismic waves emanating from distant large earthquakes. The frequency contents of triggered and ambient tremor are largely identical, suggesting that tremor does not depend directly on the nature of the source. We show here that the model of plate dynamics developed earlier by us is an appropriate tool for describing the onset of tremor. In the framework of this model, tremor is an internal response of a fault to a failure triggered by external disturbances. The model predicts generation of radiation in a frequency range defined by the fault parameters. Other specific features predicted are: the upper limit of the size of the emitting area is a few dozen km; tremor accompanies earthquakes and aseismic slip; the frequency content of tremor depends on the type of failure. The model also explains why a tremor has no clear impulsive phase, in contrast to earthquakes. A comparatively small effective normal stress (hence a high fluid pressure) is required to make the model consistent with observed tremor parameters. Our model indicates that tremor is not necessarily a superposition of low frequency earthquakes, as commonly assumed, although the latter may trigger them. The approach developed complements the conventional viewpoint which assumes that tremor reflects a frictional process with low rupture speed. Essentially our model adds the hypothesis that resonant-type oscillations exist inside a fault. This addition may change our understanding of the nature of tremor in general, and the methods of its identification and location in particular.

**Introduction**

So-called deep non-volcanic tremor (NVT) arises inside of, or in close proximity to, well-developed subduction and transform faults at a depth that is transitional between a frictionally locked region and a freely slipping region (Obara, 2002; 2009; Kostoglodov et al., 2003; Kao et al., 2005; Nadeau and Dolenc, 2005; Rubinstein et al., 2010; Peng and Gomberg, 2010). It has been observed that: (1) bursts of tremor accompany slip pulses in so-called Episodic Tremor and Slip (ETS) phenomena (e.g. Rogers, and Dragert, 2003; Obara, 2009); (2) seismic waves from either the local medium or from distant large earthquakes can trigger tremor (Obara, 2003; Rubinstein J. et al,. 2007, 2009; Peng et al., 2009; Miyazawa and Mori, 2006; Miyazawa and Brodsky, 2008; Fry et al., 2011; Zigone et al., 2012; Chao et al, 2013); (3) the intensity of tremor varies with tidal stress (Rubinstein J. et al., 2008; Nakata et al., 2008; Thomas et al., 2009; Lambert et al., 2009). While the duration and amplitude of a tremor burst vary depending on the source, the spectral composition remains essentially the same. The question arises as to how various external stress disturbances, spanning a wide range of amplitudes and frequencies, can all trigger tremor in the 1 to 30 Hz range in the fault area.

It has been shown that tremor is triggered and modulated by Rayleigh waves (Miyazawa and Mori, 2006; Miyazawa and Brodsky, 2008; Fry et al., 2011; Zigone et al., 2012), Love waves (Rubinstein et al., 2007; 2009; Peng et al., 2009; Zigone et al., 2012), as well as by P waves (Ghosh et al., 2009) and S waves (Shelly et al., 2011; Zigone et al., 2012), from distant large earthquakes. While large amplitude and a proper direction of the wave are necessary conditions for tremor triggering, these are not the only conditions required. Triggered tremor appears to be adjacent to an area of ongoing slow slip events (SSE) (Fry et al., 2011) or (in the case of short SSE) coincides with the location of the SSE source area (e.g., Hirose and Obara, 2006; Gomberg et al., 2010), suggesting that the condition that the fault be close to failure is also necessary. Triggered tremor is usually observed in the same areas as ambient tremor, with the same frequencies and polarizations (Rubinstein et al., 2009; Peng et al., 2009; Chao et al, 2013). Overall comparison of different characteristics of ambient and triggered tremor suggests that they are generated by the same physical process (e.g. Rubinstein et al., 2010).

The source of deep NVT is largely unknown. The suggested mechanisms are hydraulic fracturing (Obara, 2002; Katsumata and Kamaya, 2003; Miyazawa and Brodsky, 2008) and shear faulting (e.g. Rogers and Dragert, 2003; Shelly et al., 2007; Miyazawa and Brodsky, 2008; Nakata et al., 2011; Ben-Zion, 2012). From the time of its discovery by Obara (2002) NVT has been considered a new phenomenon since it is very distinctive from both volcanic tremor and earthquakes. Indeed the latter generate strong short impulse-like signals while NVT is a weak persistent shaking with no definite beginning and with duration from minutes to days. However, the similarity between the spectral composition of NVT and low frequency earthquakes (LFE) (Shelly et al. 2006; 2007a; La Rocca, 2009) and analysis of the relation between plate-boundary slip and LFE (Ide et al., 2007; Wech and Creager, 2007) suggest that NVT is just a swarm of LFE. Although this point of view is now dominant among geophysicists, NVT is still considered as a distinct phenomenon. At the same time, NVT features such as the absence of an impulse-like initiation, its duration (which is considerably larger than the duration of an individual LFE), persistent frequency content and an S wave type of signal, support the assumption of the resonant-type nature of tremor as a complement to the "NVT is a swarm of LFEs" hypothesis. Here we will explore this novel (resonant-type) mechanism.

Recently we developed a Frenkel-Kontorova (FK)-type model, which describes quantitatively the dynamic frictional process between two surfaces (Gershenzon et al., 2009; Gershenzon et al., 2011; Gershenzon and Bambakidis, 2013). Predictions of the model are in agreement with laboratory frictional experiments (Rubinstein S. et al., 2004; Ben-David et al., 2010). This model has also been applied to describe tremor migration patterns in ETS phenomena as well as the scaling law of slow slip events (Gershenzon et al., 2011). In the continuum limit, the FK model is described by the nonlinear sine-Gordon (SG) equation. The basic solutions of the latter are kinks and (anharmonic) lattice vibrations (phonons) (e.g. McLauglin and Scott, 1978) which, in our context, may be interpreted as slip pulses and radiation respectively. In the framework of the model, radiation may arise due to a variety of mechanisms such as acceleration or deceleration of a slip pulse, interaction of a slip pulse with large asperities, and the action of an external stress disturbance on the frictional interface. The first two mechanisms may be used to describe generation of tremor during ETS events and will be considered in detail in a future publication. In this article we will focus on the latter mechanism.

Here is our suggested scenario (see Figure 1). The low frequency seismic wave generated by a distant earthquake increases the tangential stress and/or decreases the effective normal stress in the vicinity of a fault, so the Coulomb stress temporarily increases and static friction decreases (Hill, 2012). There are always spots within a fault with residual tangential stress. Such spots may remain, for example, after a slip pulse passes the region. Thus, a seismic wave with sufficiently large amplitude and proper direction may trigger local failure (slip), exciting a radiation mode inside the fault. This radiation (as a small-amplitude, localized relative motion of plate surfaces with zero net slip) propagates along the fault attenuated due to friction and geometrical spreading. Since the fault is immersed in a 3D solid body, the radiation inside the fault will generate S waves (tremor) propagating as far as the Earth's surface. It is important to note that the frequency of these waves is defined by the radiation frequency, hence by the fault parameters, and does not depend on the frequency of the external source. Thus, in the model we propose, tremor is a result of specific resonant-type radiation generated inside a fault. This radiation may be initiated by LFEs as well as by other types of local failure such as regular earthquakes, very low frequency (VLF) earthquakes and aseismic slip.

The rest of the manuscript is organized as follows. Section "The Model" describes the basics of the model. Further development of the model and detailed description of the spatial and temporal distribution of the radiation mode are presented in section "Non-Volcanic Tremor". The results are applied to the quantitative assessment of tremor parameters (section

"Discussion"). In the conclusion we summarize the specific prediction of our model and discuss the possibilities for its examination.

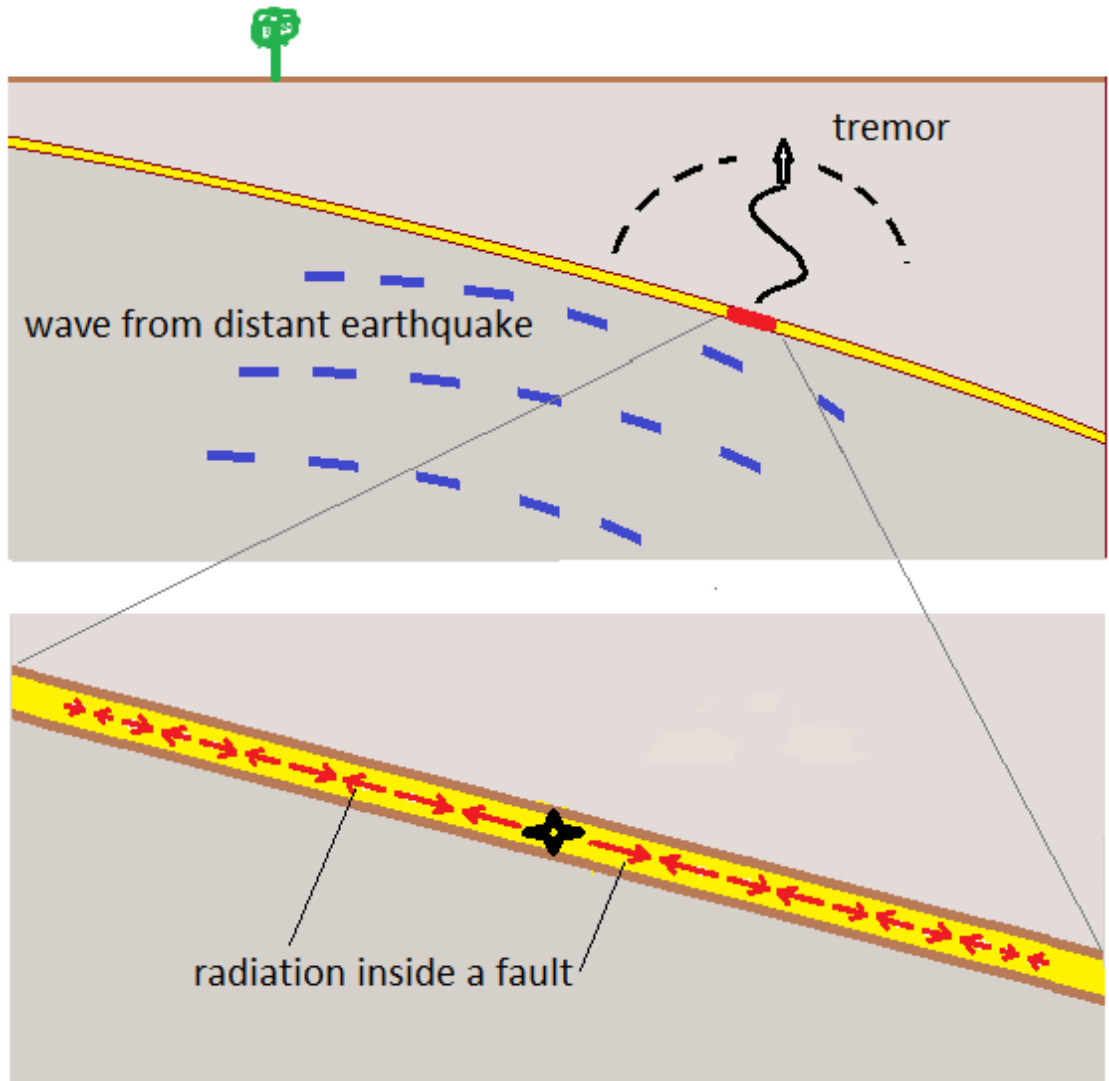

**Figure 1**. Scenario for tremor generation. The initial failure (black object at the bottom panel), triggered by a seismic wave of a distant earthquake, produces oscillations (shown by red arrows) propagating along a subduction fault (in yellow). These oscillations generate an S-type seismic wave (tremor) propagating to the surface.

**The Model**

It has been shown that the dynamics of a frictional surface may be described by the FK model (Gershenzon et al., 2009; Gershenzon and Bambakidis, 2013). Here we provide a brief description of the model and its application to slip dynamics. We will consider the asperities on one of the frictional surfaces (together with the surrounding material, Figure 2a) as forming a linear chain of balls of mass $M$, each ball interacting with its nearest neighbors on either side via spring forces of constant $K_b$ (Figures 2b and 2c). The asperities on the opposite frictional surface will be regarded as forming a rigid substrate which interacts with the masses $M$ via a periodic potential. Then we can apply the one-dimensional FK model to describe the slip dynamics (Frenkel and Kontorova, 1938; Hirth and Lothe, 1982):

$$M\frac{\partial^2 u_i}{\partial t^2} - K_b(u_{i+1} - 2u_i + u_{i-1}) + F_d \sin(\frac{2\pi}{b}u_i) = F(x,t) - f_{fr}(x,t,\frac{\partial u_i}{\partial t}), \qquad (1)$$

where $u_i$ is the shift of ball $i$ relative to its equilibrium position, $b$ is a typical distance between asperities, $t$ is time, $F_d$ is the amplitude of the periodic force on the mass $M$ associated with the periodic substrate potential, $f_{fr}$ is the frictional (or dissipative) force, and $F$ is the external (or driving) force. Using the continuum limit approximation (the latter can be used if the typical distance between asperities is much less than the typical wave length of the processes considered) and expressing the coefficients $M$, $K_b$ and $F_d$ through the elastic parameters and the normal stress between frictional surfaces, we may express equation (1) in the form (Gershenzon and Bambakidis, 2013),

$$\frac{\partial^2(\frac{2\pi u}{b})}{\partial(tc/b)^2} - \frac{\partial^2(\frac{2\pi u}{b})}{\partial(x/b)^2} + A^2 \sin\left(\frac{2\pi u}{b}\right) = (F - f_{fr})\frac{2\pi A^2}{\mu b^2}, \qquad (2)$$

where $c^2 = \frac{2\mu}{\rho(1-\nu)} = \frac{c_l^2(1-2\nu)}{(1-\nu)^2}$, $c_l$ is the longitudinal acoustic velocity (or P wave velocity); $\mu$ is the shear modulus and $\nu$ is the Poisson ratio. Note that $c_s < c < c_l$, where $c_s$ is the shear wave velocity (or S wave velocity). The dimensionless parameter $A$ reflects how deeply the asperities from two opposing surfaces interpenetrate, and its value can be considered as the ratio between actual and nominal contact areas, hence as the ratio between the effective normal stress $\Sigma_N$ and the penetration hardness $\sigma_p$: $A \approx \Sigma_N/\sigma_p$. The equivalent form of equation (2) is the well-known perturbed sine-Gordon (SG) equation:

$$\frac{\partial^2 u}{\partial t^2} - \frac{\partial^2 u}{\partial x^2} + \sin(u) = \Sigma_S^0 - f, \qquad (3)$$

where the dimensionless variables $u$, $x$ and $t$, respectively, are in units of $b/(2\pi)$, $b/A$ and $b/(cA)$, and the source terms $\Sigma_S^0$ and $f$ are the external shear stress and frictional force per unit area, both in units of $\mu A/(2\pi)$. The variables $\varepsilon = \sigma_s = \frac{du}{dx}$ and $w = \frac{du}{dt}$ are interpreted as the dimensionless strain, stress and slip velocity in units of $A/(2\pi)$, $\mu A/\pi$ and $cA/(2\pi)$, respectively.

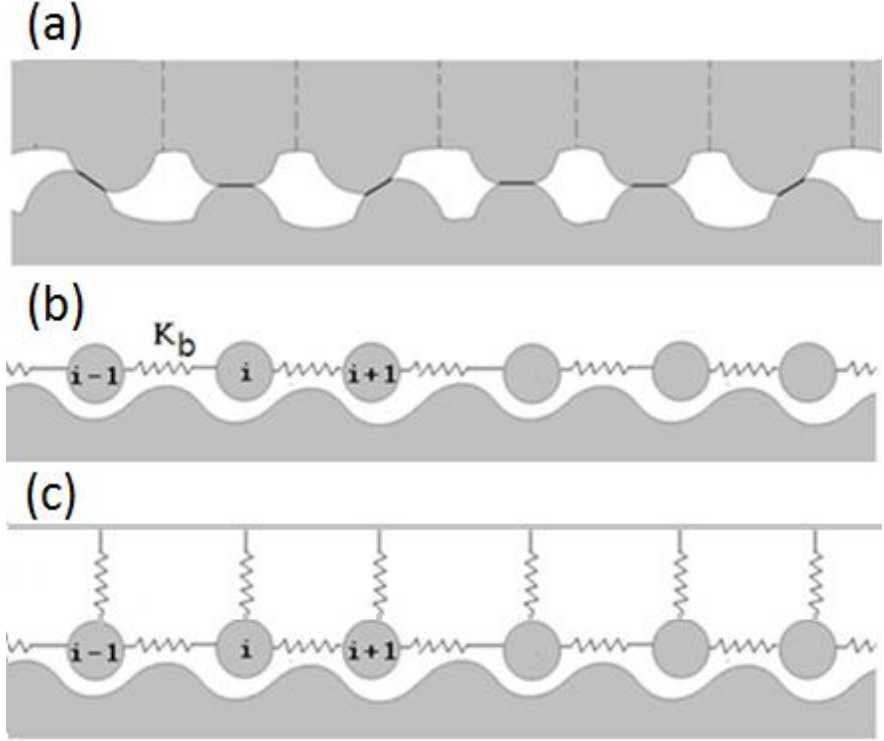

Figure 2. Schematic of asperity contact (a) and chain of masses interacting via elastic springs and placed in a periodic potential (substrate) (b) and (c). The balls represent asperities. The sine-shaped surface is the opposite plate. The horizontal and vertical harmonic springs model interaction between asperities on the same and opposite plates, respectively. In the classic FK model the harmonic forces arise due to motion of a ball along the uneven surface in a gravitational field (b), whereas in our model the harmonic forces arise due to the vertical springs (c). The mathematical descriptions of models (b) and (c) are identical.

The basic solutions of the SG equation are kinks (solitons), breathers and anharmonic vibrations (phonons). In a previous article it was shown that the kink solution can be used to model the dynamics of a slip pulse in ETS events (Gershenzon et al., 2011). In the next section we will consider the phonon solution to describe non-volcanic tremor. First we will discuss the solution of the equation (3), which has the form of a simple traveling wave $u = u(x - Ut) = u(\xi)$ propagating with dimensionless velocity $U$, with $U^2 > 1$. The solution is:

$$u = 2\arcsin(m^{\frac{1}{2}} sn[\frac{\xi - \xi_0}{(U^2-1)^{\frac{1}{2}}}, m]), \quad (4)$$

where $sn$ is the Jacobi elliptic function of modulus $m$ ($0 \leq m \leq 1$) and $\xi_0$ is the initial phase. The value of $m$ is defined in terms of the wave amplitude $a$ by the relation: $m = [(1 - \cos(a)/2)]^{0.5}$. The nonlinear dispersion equation for the "phonon" mode is (McLauglin and Scott, 1978):

$$\omega^2 - k^2 = \frac{\pi^2}{4K^2(m)}, \tag{5}$$

where ω is the angular frequency in units of *cA/b*, *k* is the wave number in units of *A/b*, and *K* is the complete elliptic integral of the first kind. Figure 3 show the dependence of ω on *k* for various values of *m* (and hence wave amplitude). One can see that if $m_1 \equiv 1-m \ll 1$ the wave amplitude is large ($a \approx \pi$) and the right hand side of equation (5) is very small, thus ω≈*k* (the bottom curve on Figure 3). If the wave amplitude is small, i.e. *m*<<1, the dispersion relation (5) is simplified (the top curve in Figure 3):

$$\omega^2 - k^2 = 1. \tag{6}$$

So for waves with amplitude much less than *b* (dimensionless amplitude *a*<<π) the dimensionless group velocity *V* of the wave packet *V* (in units of *c*) is:

$$V = \frac{d\omega}{dk} = \frac{k}{(1+k^2)^{0.5}}. \tag{7}$$

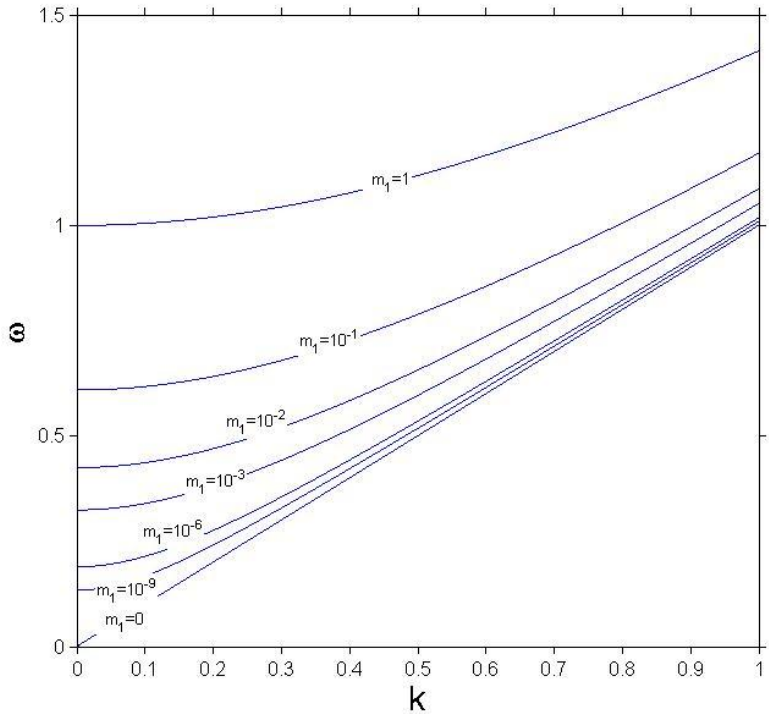

Figure 3. The nonlinear dispersion equation for the radiation mode: the dependence of angular frequency ω on wave number *k* for different values of *m*, hence, wave amplitude.

Two important conclusions follow from equations (6) and (7): (1) the angular frequency cannot be less than unity ($\omega \geq 1$), which means that the minimal frequency of the disturbance generated inside a fault, $f_{min} = \left(\frac{cA}{b}\right)\left(\frac{\omega}{2\pi}\right) = \left(\frac{cA}{2\pi b}\right)$ in dimensional units, is defined by the fault parameters and effective normal stress only and does not depend on the parameters of a particular source; (2) the group velocity of a disturbance propagation along a fault may vary from values much less than $c$ (if $k^2 \ll 1$ then $V \approx k \ll 1$) to the value $c$ (if $k^2 \gg 1$ then $V \approx 1$).

**Non-Volcanic Tremor**

Let us assume that there is an area on the fault close to failure, i.e. an area with residual localized shear stress almost equal to the value of the static frictional force per unit area. This island of residual stress may have remained after passage of a slip pulse (for example, after an ETS event) due to the presence of an asperity with size larger than the typical size of surrounding asperities. Suppose that a seismic wave of large amplitude arrives at this spot, increasing the tangential stress and/or decreasing the normal stress in such a way that locally and temporally the tangential stress exceeds static friction. The ensuing failure (slip) overcomes or destroys this large asperity, producing a regular or slow earthquake or simply aseismic slip. The stress impulse can be considered as a source term for the perturbed SG equation, and in this section we shall see how this can generate tremor. We will consider two examples of an initial disturbance, modeling 1) fast failure such as a regular earthquake or LFE, and 2) slow failure such as VLF earthquake or aseismic slip.

*Fast initial failure*

Let us model the spontaneous failure within a fault as an impulse localized in space and time. Using this as the source term in equation (3), i.e. $\Sigma_S^0 - f = a_{init}\delta(x)\delta(t)$, where δ is the Dirac delta-function and $a_{init}$ is the strength of the disturbance, we look for the "phonon" radiation field inside the fault produced by this impulse. Results of a perturbation analysis of the SG equation (McLauglin and Scott, 1978) gives the "slip field" produced by the perturbation (see Appendix A for details):

$$u(x,t) = \frac{a_{init}}{2\pi} \int_{-\infty}^{\infty} \frac{\sin(\varphi t)\exp(-i\phi x)}{\varphi} d\phi. \qquad (8a)$$

where $\varphi = (\phi^2 + 1)^{0.5}$. The associated spatial and temporal distribution of slip velocity $w$ and shear stress σ are

$$w(x,t) \equiv \frac{du}{dt} = \frac{a_{init}}{2\pi} \int_{-\infty}^{\infty} \cos(\varphi t) \exp(-i\phi x) d\phi, \qquad (8b)$$

$$\sigma(x,t) \equiv \frac{du}{dx} = \frac{-a_{init}}{2\pi} \int_{-\infty}^{\infty} \frac{i\phi}{\varphi} \sin(\varphi t) \exp(-i\phi x) d\phi. \qquad (8c)$$

Figure 4 shows the results of a numerical integration of equation (8a) for $a_{init} = 1$. One can see that the disturbance (as a small relative shift of plates) originating at point $x$=0 at time $t$=0 propagates along a fault in both directions with unit velocity (velocity $c$ in dimensional units) (Figure 4 a-d). The wave number $k$ ranges from large values close to the wave fronts to small values close to the center. The value of $k$ at the center decreases in time and becomes much less than unity when $t$>>$2\pi$. In this case $\omega$≈1 (see equation (6) and Figure 4e), so after a short time the frequency of the oscillation in close proximity to the center reaches the value $f$=$\omega$/($2\pi$)≈1/($2\pi$) and does not change much thereafter. The oscillation frequency at points close to the fronts is higher (see Figure 4 f-h). Therefore the angular frequency ranges from a low of $\omega$=1 at the center to higher frequencies $\omega = (1 + k_{init}^2)^{0.5}$ at the periphery of the disturbance, where $k_{init}$ is the characteristic wave number of the initial disturbance (see Figures 4 e-h). The results of the calculation of the spatial distribution of $w$ and $\sigma$ for different times are presented in Figure 5.

Let us solve equation (3) with the right hand side of the form:

$$\Sigma_S^0 - f = a_{init} \left(\frac{9}{\pi}\right)^{0.5} exp(-9x^2)\delta(t) + \sigma_0 - [\alpha_s sign(u_t) + \alpha_d u_t], \qquad (9)$$

where the first term on the right hand side represents the initial conditions that the integral over $x$ equal $a_{init}$, the second term is the external shear stress, and the last term denotes friction (which includes both static and dynamic terms). The boundary conditions are $u(x=x_-)=u(x=x_+)=0$ and $u_t(x=x_-)=u_t(x=x_+)=0$, where $x_-$ and $x_+$ are the left and right positions of the boundary. Figures 6 and 7 show the evolution of a signal in time and space for symmetric and non-symmetric boundary conditions, respectively (relative to the position of initial source). As one can see, the disturbance is propagated to the boundary and then reflected back, gradually decreasing. Note that the presence of nonzero external stress just shifts the equilibrium position by a certain distance smaller than $b$ and practically no effect on the oscillations if $\sigma_0 \ll 1$, which is almost always the case. So in the remainder of the paper we will simply assume $\sigma_0 = 0$.

Thus, we have showed that a localized disturbance (a source) may initiate oscillations which spread inside a fault in both directions along $x$. The oscillatory area is limited by two fronts propagating with velocity $c$ (in dimensional units). Obviously within a few seconds or less the fronts will leave the area of applicability of our model. Indeed if $c$=5 km/s and $t$=10s the size of the disturbed area is about 100 km. The frictional behavior along a fault due to temperature and water pressure variations presumably changes on a smaller scale. By the time a signal reaches the boundary it may already have disappeared due to friction. If this is not the case we must introduce boundary conditions.

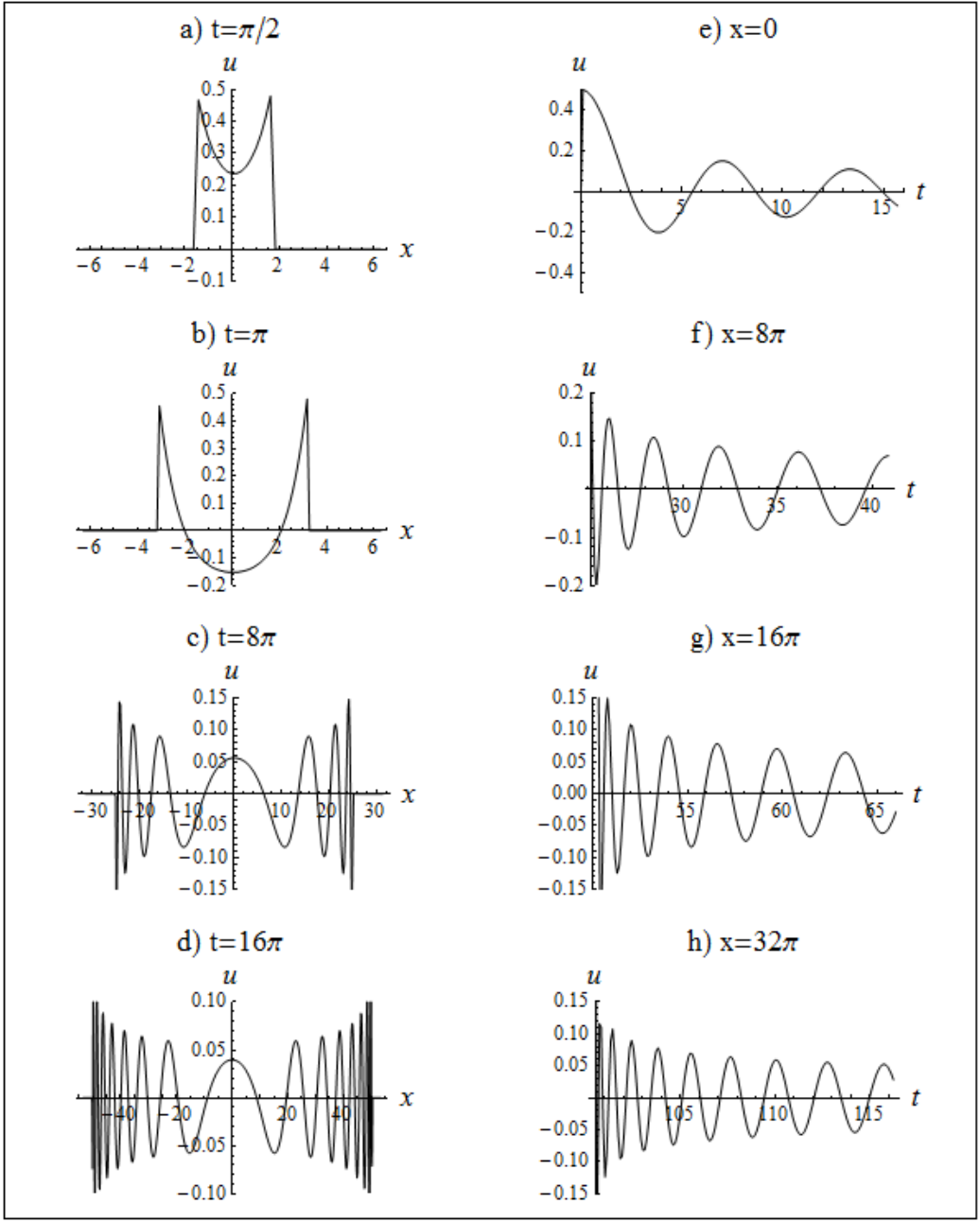

**Figure 4**. The evolution of the shift disturbance in space (along a fault) for various times (panels on the left) and in time for various distances from the center (panels on the right). Disturbance originates at point $x=0$ and time $t=0$ due to the external source $\delta(x)\cdot\delta(t)$ and propagates in both directions. The local wave number at any particular time decreases going from the center to the wave fronts (Figure 4c and 4d). The period of oscillation at point $x=0$ approaches $2\pi$ ($f\approx 1/(2\pi)$) after a short time ($t>2\pi$) from the beginning (see Figure 4e). The oscillation period at points $x=8\pi$, $16\pi$, $32\pi$ progressively decreases compared with the value at point $x=0$ (see Figure 4 f-h). The disturbance at the center plays the dominant role in the frequency distribution of the emitted tremor; the peripheral disturbance contributes to its high frequency range.

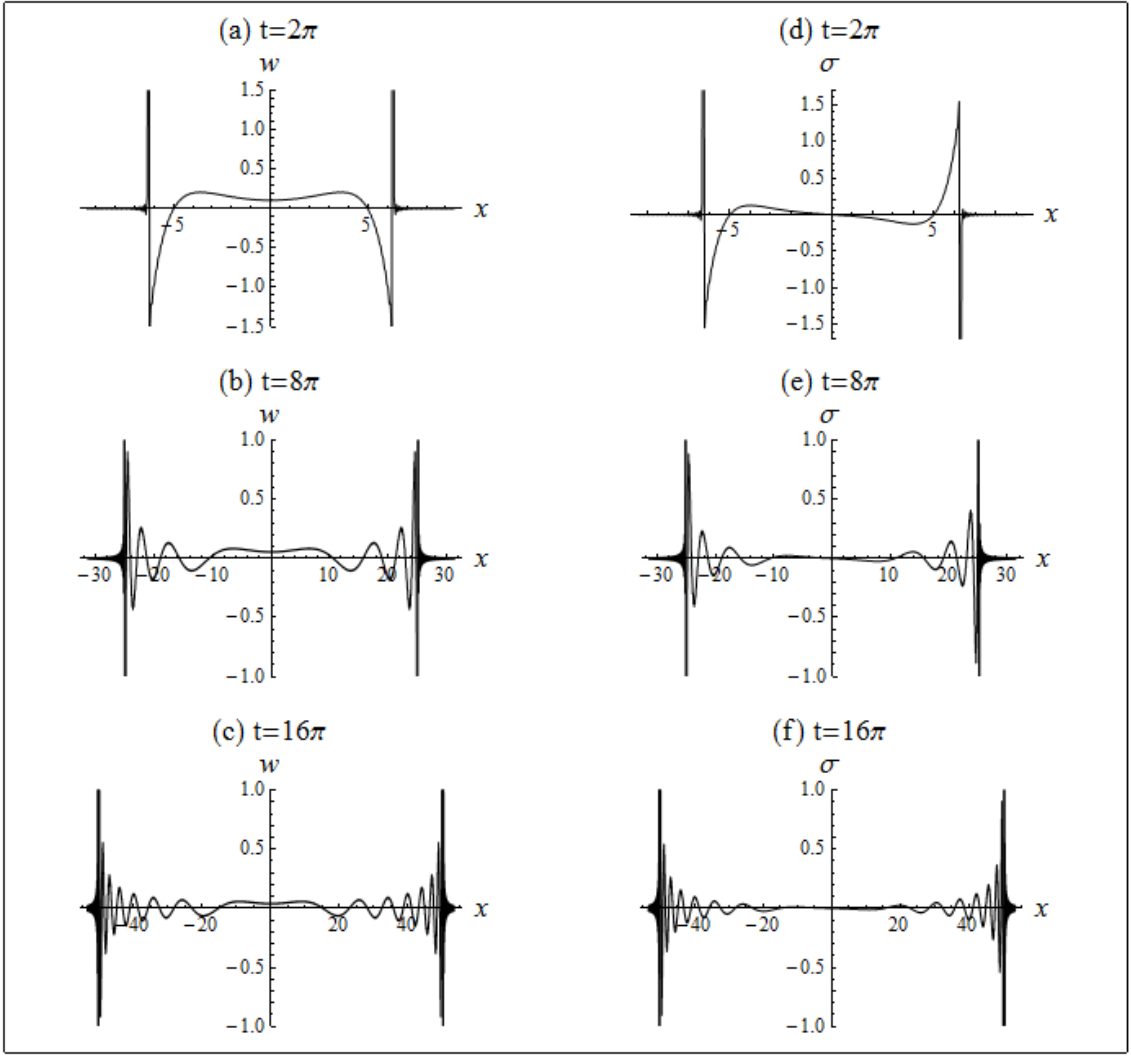

**Figure 5.** Spatial distribution of slip velocity $w$ (panels on the left) and shear stress $\sigma$ (panels on the right) along a fault for various times. Disturbance originates at point $x=0$ and time $t=0$ by the external source $\delta(x) \cdot \delta(t)$ and propagates in both directions. The high frequency oscillations seen on all figures are an artifact of the calculation.

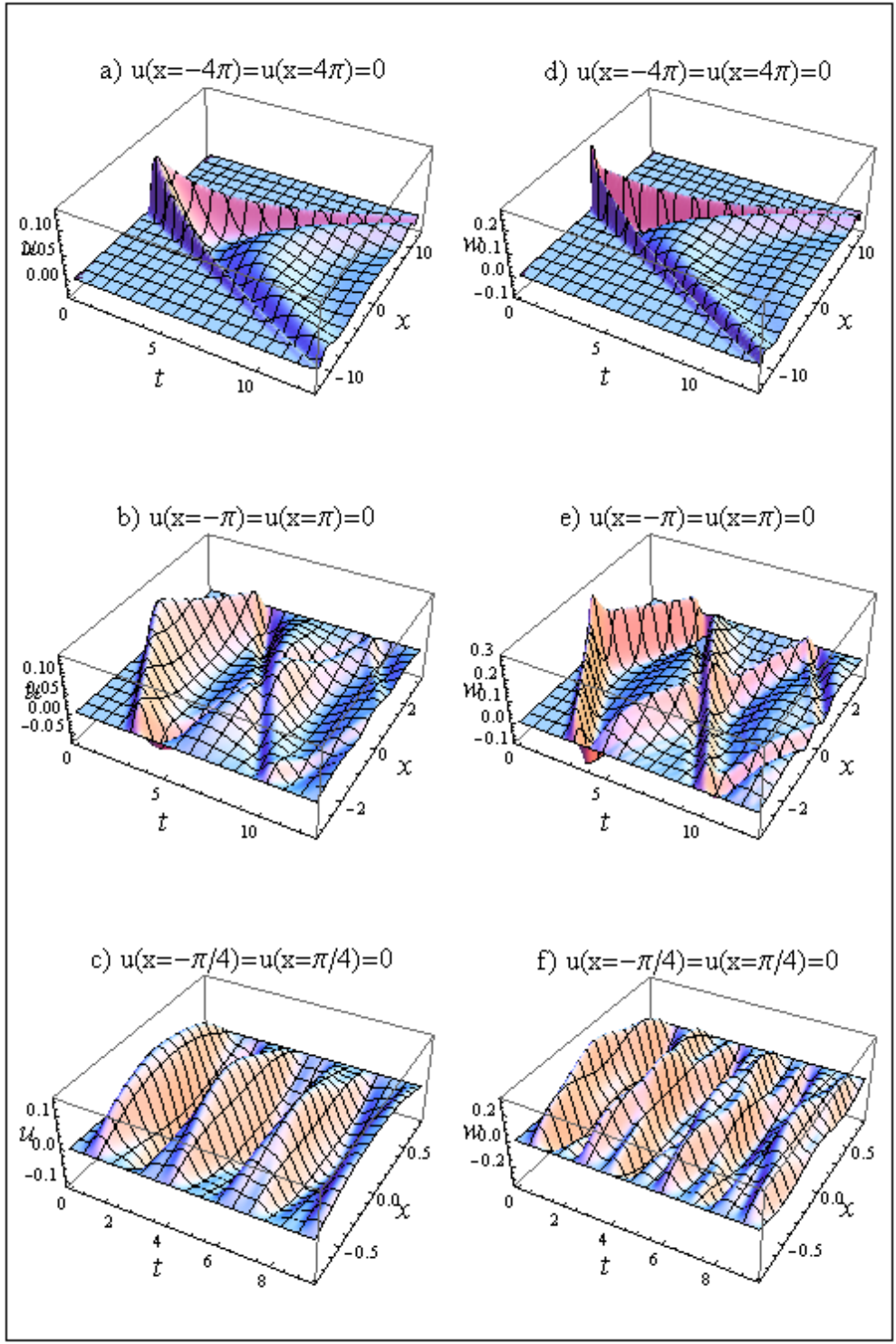

**Figure 6.** Temporal and spatial evolution of a signal (slip *u* and slip velocity *w*) inside a fault for different values of boundary position, computed for the fast initial failure, i.e. for LFE event. Computation made with $\alpha_s = \alpha_d = 0.025$.

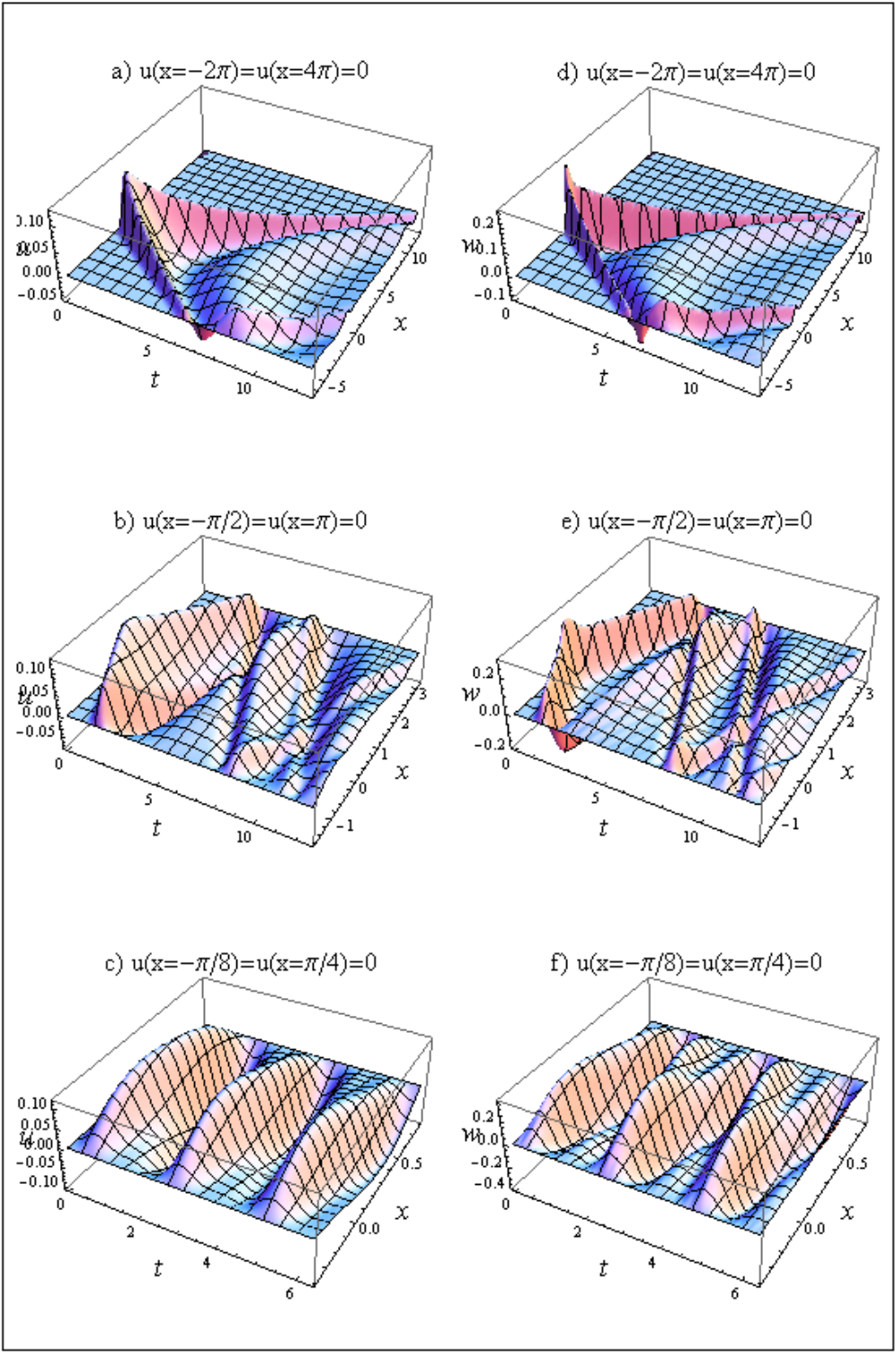

**Figure 7.** The same as in Figure 6 for non-symmetrically placed boundary positions.

Since the plate surface is part of the earth's crust, its periodic localized oscillations will generate *S*-type seismic waves with the same mix of frequencies; the latter may propagate through the crust to the Earth's surface. Thus the model predicts the generation of tremor in the ω range from 1 to $(1 + k_{init}^2)^{0.5}$. Oscillations close to the center contribute to the lower portion of the tremor frequency range while oscillations at the periphery contribute to the higher portion. To determine the disturbances (tremor) at a given point $(x_s, y_s, z_s)$ on the Earth's surface produced by the disturbance generated inside a fault, one needs to integrate velocity *w* over the entire source. In the far zone approximation the expression for calculation of Earth's surface shift $u_s$ is (Aki & Richards, 1980):

$$u_s(x_s, y_s, z_s, t) = \int_0^t \int_{x_-}^{x_+} \int_{y_-}^{y_+} \frac{gw(x, t' - r/c_s)}{r} dx dy \, dt', \tag{10}$$

where $y_-$ and $y_+$ form the boundary of the source in the *y* direction *r* is the distance between source and observation point and *g* is a coefficient. Since our model is one-dimensional, we will assume that the dimension of the source in the *y* direction is constant. Inserting the solution of equation (3) into the equation (10) we find $u_s$ after integration over the *x* and *y* coordinates and time *t'*. In the calculation we assume for simplicity that the effective size of a disturbance is much smaller than the distance between the source and the measurement point on the Earth's surface. Figure 8 show the results of a numerical integration of the velocity $v = du_s/dt$ and spectral content for various values of friction coefficients. Increasing the friction coefficients decreases (obviously) the signal life-time and increases the spectral amplitude at higher frequencies. The dashed lines show the spectral density of the same signal but including a frequency-dependent attenuation during propagation from the source to the measurement point. To calculate the attenuation we multiply the spectrum by exp(−π*ft*/*Q*), where the travel time *t* = 15s and *Q* = 350 (using the same approach as Beroza and Ide (2011)). Figure 9 depicts the results of calculations with various friction laws, i.e. $\alpha_d u_t$, $\alpha_s sign(u_t)$ and $\alpha_s sign(u_t) + \alpha_d u_t$. As one can see, the decay of the signal is larger for static friction than for dynamic friction (compare Figures 9a and b).

The boundary conditions obviously affect the shape and frequency content of a signal (tremor); however the main characteristics, i.e. central frequency and spectral fall-off law, remain practically unchanged when boundaries are placed outside the range $-2\pi < x < 2\pi$. However, if the boundaries are inside this range and the boundary conditions are $u(x_-) = u(x_+) = 0$ (no movement) the maximum in the spectral density shifts to a higher frequency (see Figure 10). In the case of the "free movement" boundary conditions, $u_x(x_-) = u_x(x_+) = 0$, the frequency content is practically the same as for the case with no boundary conditions (not shown).

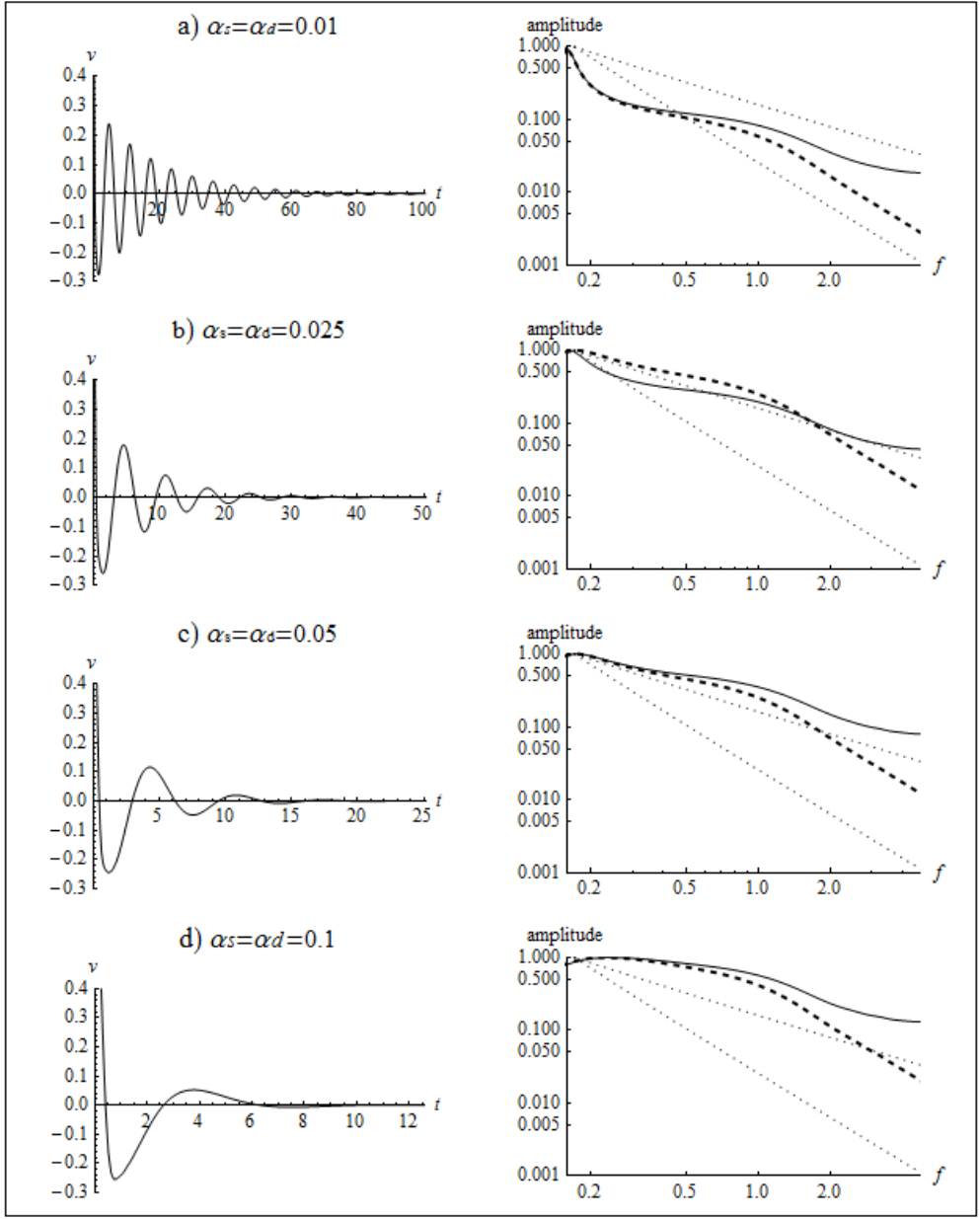

**Figure 8**. The calculated tremor, i.e. velocity of Earth's surface movement (in arbitrary units), versus time (in dimensionless units) produced by the "fast" disturbances for various values of friction coefficients. Also shown is the corresponding frequency content of tremor in the range from $1/(2\pi)$ to $30/(2\pi)$ (hence from 1 to 30 Hz in dimensional units). The reference fall-offs (dotted line) are $f^{-1}$ and $f^{-2}$. The dashed lines show the spectral density of the same signal but taking into account frequency-dependent attenuation during its propagation from the source to the measurement point. To calculate attenuation we multiply the spectrum by $\exp(-\pi f t/Q)$, where the travel time $t = 15$ s and $Q = 350$ (Beroza and Ide, 2011).

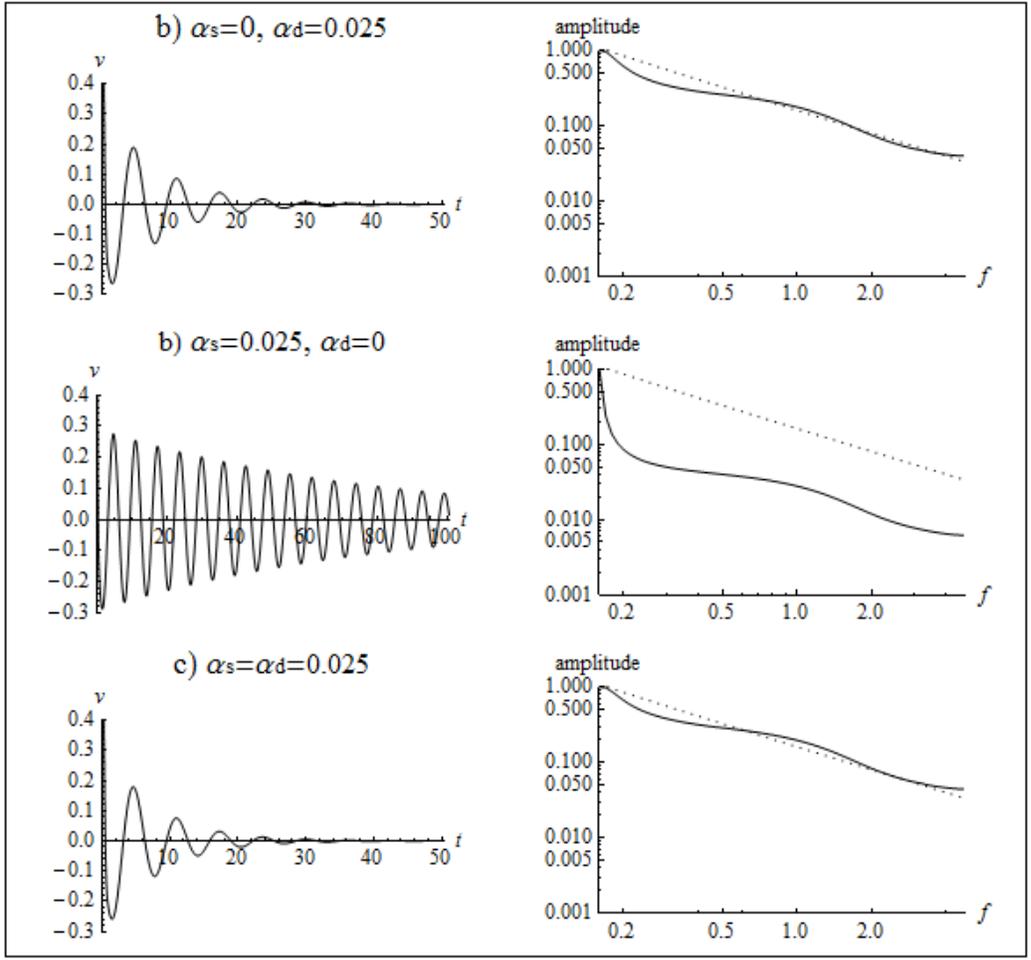

**Figure 9**. The same as Figure 8 for the three friction laws: $\alpha_d u_t$, $\alpha_s sign(u_t)$ and $\alpha_s sign(u_t) + \alpha_d u_t$ . The reference fall-off (dotted line) is $f^{-1}$.

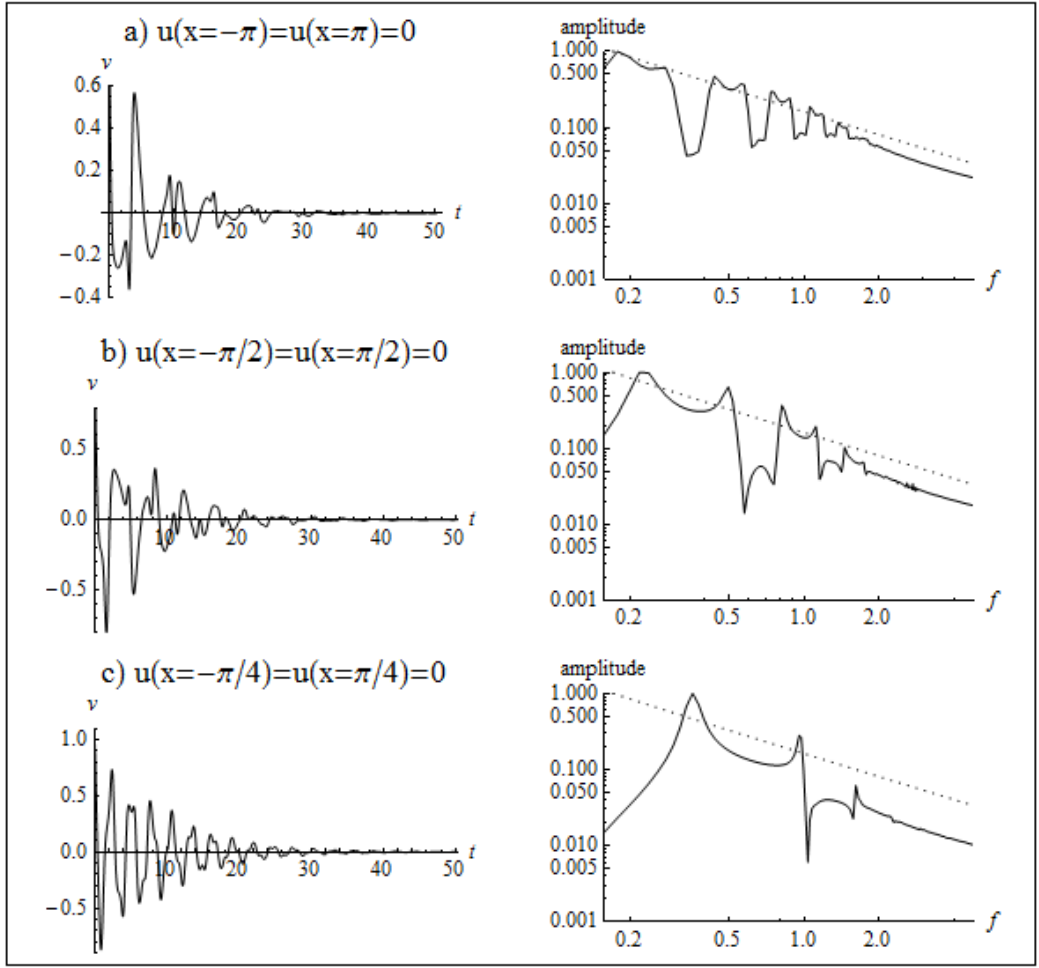

**Figure 10**. The calculated tremor produced by the "fast" disturbances for various placements of boundaries. Also shown is the corresponding frequency content of tremor. The reference fall-off (dotted line) is $f^{-1}$. Computation made with $\alpha_s = \alpha_d = 0.025$.

*Slow initial failure*

Now consider the case where the source term in equation (3) has the form $\Sigma_S^0 - f = a_{init}\delta(x)\eta(t)$, where η is the Heaviside function. As before, perturbation analysis gives (see Appendix A equation (A3)):

$$u(x,t) = \frac{a_{init}}{2\pi}\int_{-\infty}^{\infty}\frac{(1-\cos(\varphi t))\exp(-i\phi x)}{\varphi^2}d\phi \tag{11a}$$

for the radiation field, and

$$w(x,t) = \frac{a_{init}}{2\pi}\int_{-\infty}^{\infty}\frac{\sin(\varphi t)\exp(-i\phi x)}{\varphi}d\phi \tag{11b}$$

and

$$\sigma(x,t) = \frac{-a_{init}}{2\pi} \int_{-\infty}^{\infty} \frac{i\phi}{\varphi}(1 - \cos(\varphi t)) \exp(-i\phi x) \, d\phi \tag{11c}$$

for the associated spatial and temporal distributions of slip velocity and shear strain, respectively. Figures 11 and 12 show the results of a numerical integration of equations (11(a-c)). From these figures one can see that, although the shape of the oscillations is different from the case of fast initial failure (compare Figures 4 with 11 and Figures 5 with 12), the main features are the same, namely 1) fronts propagate with unit velocity, 2) the disturbance includes small $k$ at the center and large $k$ close to the boundaries, and 3) the frequency of oscillation in the center area is close to $2\pi$.

Solving equation (3) numerically with the right hand side in the form:

$$\Sigma_S^0 - f = a_{init}\left(\frac{3}{\pi}\right)^{0.5} exp(-3x^2)\eta(t) - \alpha_s sign(u_t) + \alpha_d u_t, \tag{12}$$

we find the spatial and temporal evolution of the signal (Figure 13). Inserting the solution obtained to equation (10) we compute the velocity of the Earth's surface movement and its spectral composition (Figure 14). Note that, although the central frequency of tremor initiated by the fast and slow failures are the same, the spectral compositions is noticeably different, i.e. spectral fall-off is larger for the tremor initiated by slow initial failure for the same value of friction coefficients (see Figure 8 and 14).

**Discussion**

Our model contains two adjustable parameters: the typical distance $b$ between asperities and the dimensionless parameter $A$. In a previous article (Gershenzon et al., 2011) we showed that a slip pulse in an ETS event could be represented as a solitonic solution of equation (3). Then the parameter $b$ should be the typical slip produced by one ETS event, i.e. $b \approx 30$ mm (Rogers and Dragert, 2003). Note that the distance $b$ is not the same as the size of an asperity the failure of which initiates (triggers) tremor. The parameter $A$ is the ratio between the effective normal stress and the penetration hardness (Gershenzon and Bambakidis, 2013), both of which are unknown in our case. We can estimate $A$ assuming that the predicted minimal frequency of tremor, $f = cA/(2\pi b)$ in dimensional units, corresponds to the lowest frequency of the observed velocity spectrum of tremor, ~1 Hz. Thus we find that $A \approx 4 \cdot 10^{-5}$ if $f$=1 Hz, $b$=30 mm and $c$=5 km/s. Having the values of these two adjustable parameters as determined from observed well-defined quantities, we now use our model to calculate the values of other parameters which are difficult to measure.

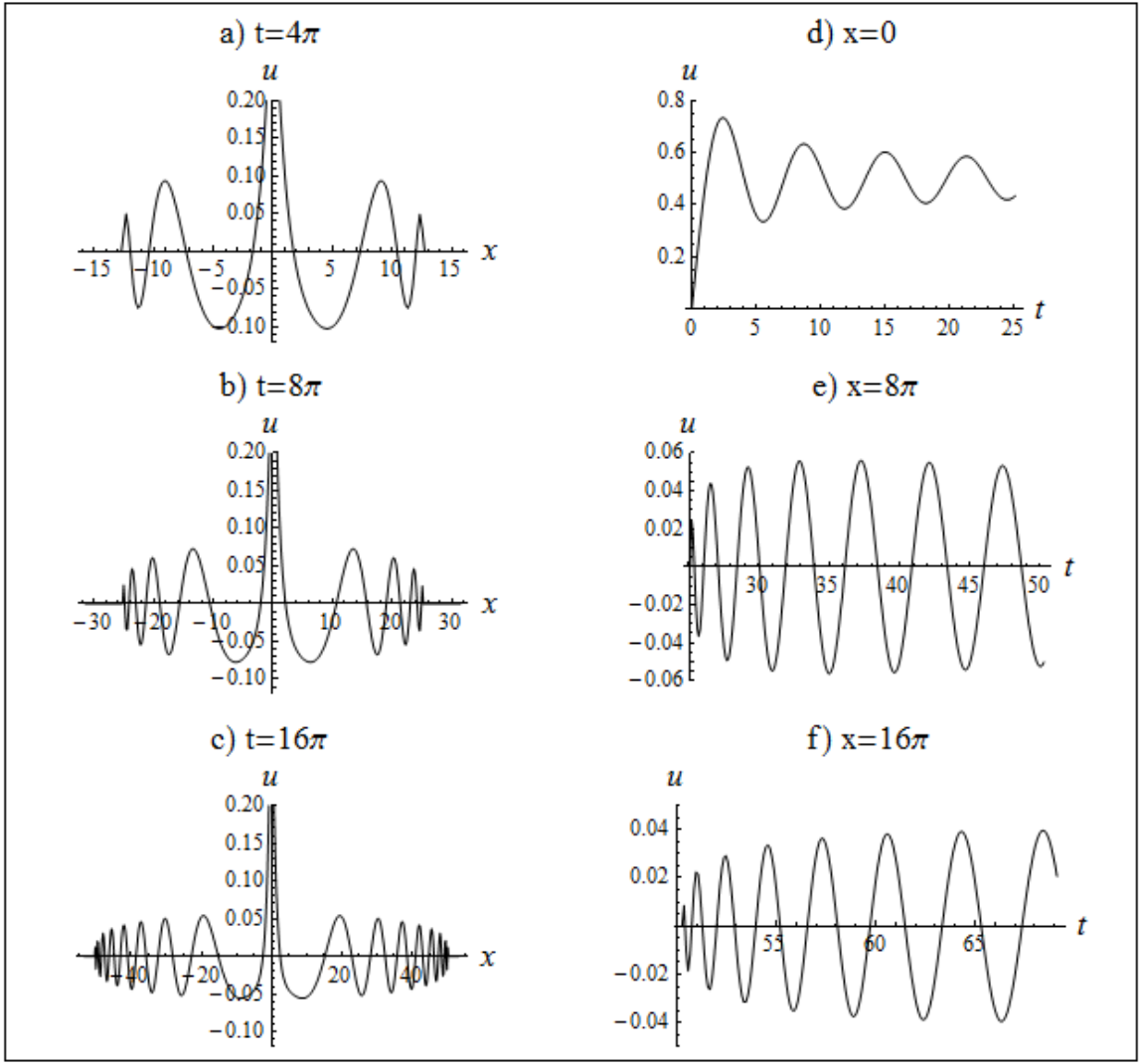

**Figure 11**. The same as in Figure 4 for the external source $\delta(x)\cdot\eta(t)$.

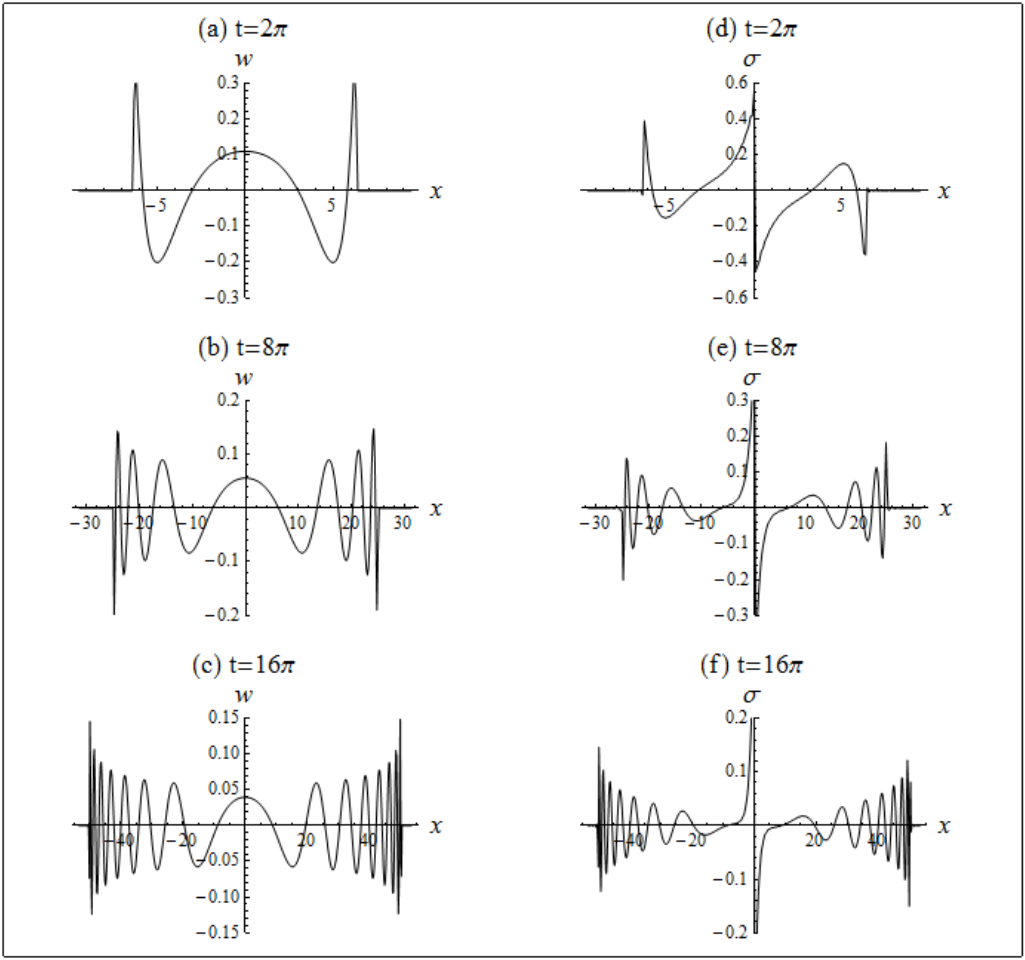

**Figure 12**. The same as in Figure 5 for the external source δ(*x*)·η(*t*).

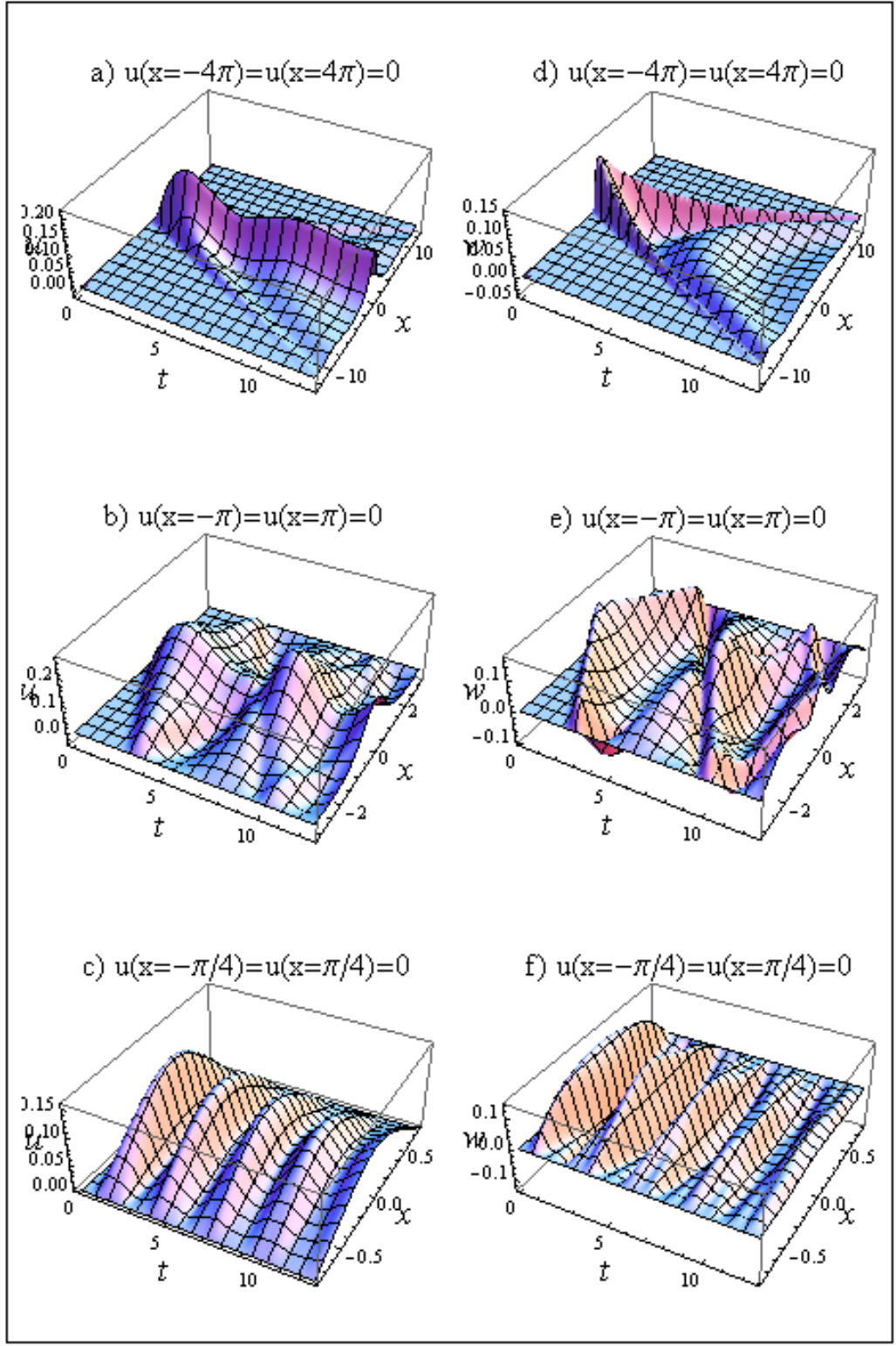

**Figure 13**. The same as in Figure 7 for slow initial failure, i.e. for VLF earthquakes as the tremor source.

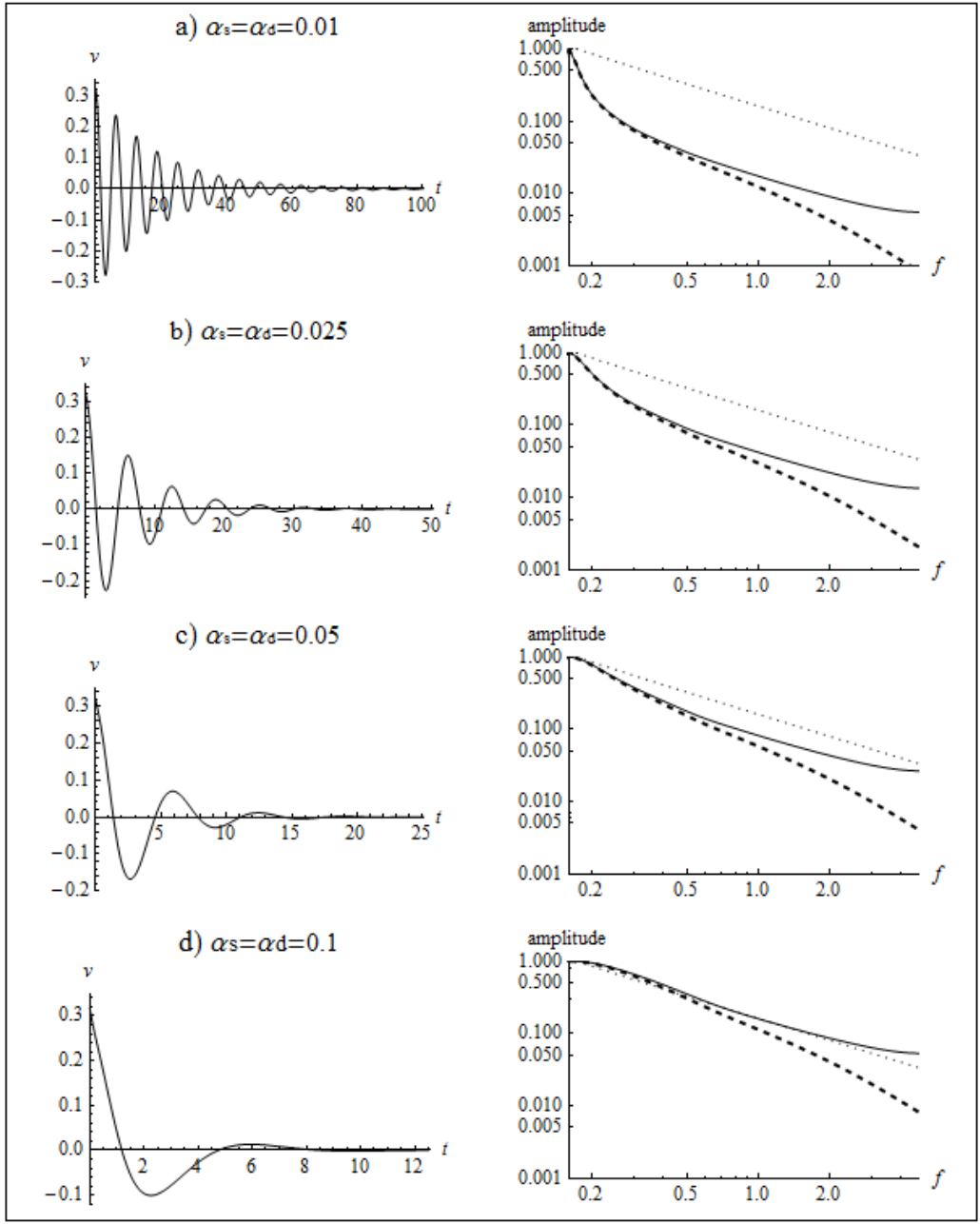

**Figure 14**. The same as in Figure 8 for slow initial failure.

Supposing that $\sigma_p \approx 0.018\mu(1+\nu)$ (Rabinowicz, 1965) and taking $\mu=30$ GPa and $\nu=0.3$, we estimate the value of the effective normal stress, $\Sigma_N \approx A \cdot \sigma_p$, to be 25 kPa. Although this estimate is not particularly accurate due to the uncertainty of the $\sigma_p$ value, it nevertheless allows us to conclude that the effective normal stress is quite low, i.e. a high fluid pressure is required for the model to work. Using the failure slip $a_{init}$, which initiate radiation, along with the results of our calculations (Figures 4 and 5) we may estimate the radiation parameters inside a fault, i.e. the maximal amplitudes of the slip $a_{rad}$ (not the initial failure slip, but a slip produced by radiation), the slip velocity $a_w$ and the shear stress $a_\sigma$. Supposing that $a_{init}$ = 10 mm we find $a_{rad} \approx$ (0.05-0.1) · $a_{init}$ = (0.5-1) mm (see Figures 4 and 11). To estimate $a_w$ and $a_\sigma$ we need to calculate the value of $a_{init}$ in dimensionless units, which is $a_{init}$ = $(2\pi/b)$ · (10mm) $\approx$ 2.09. Now we can find $a_w \approx (cA/2\pi)$ · (0.05-0.1) · $a_{init} \approx$ (2-4) mm/s and $a_\sigma \approx (\mu A/\pi)$ · (0.05-0.1) · $a_{init} \approx$ (40-80) kPa, respectively (see Figures 5 and 12).

Occurrence of radiation inside a fault is a key to the model considered. The question arises about the conditions required for the existence of such oscillatory relative movement (slip) of the frictional surfaces. Could static friction prevent its appearance? First of all note that the process we have considered does not include "conventional friction". In the conventional frictional process it is implicitly assumed that slip exceeds at least one inter-asperity spacing $b$. In our case the slip due to oscillations is smaller (by definition) than $b$. From the mathematical point of view, if slip is larger than $b$ the disturbance is a kink (or soliton), not a phonon. Here we consider the phonon mode only. So friction during oscillations with amplitude smaller than the inter-asperity size is not the same as (and supposedly much smaller than) conventional friction. Furthermore we have already shown that the effective normal stress, and thus the frictional force, should be extremely low ($\Sigma_N \approx$ 25 kPa) for the model to work. We have estimated above that the amplitude of the shear stress in oscillations is of the same order as the effective normal stress, which means that friction should not suppress the oscillations. Under these conditions it is reasonable to assume that radiation initiated by a failure inside the fault can exist. Note that Ben-Zion (2012) has proposed the existence of a "critical (near-) zero weakening" during slip below the seismogenic zone, to describe the appearance of NVT. This assumption is related, but not identical, to ours.

In the scenario proposed, the radiation gradually increases with time. The typical time of tremor development is $2\pi$ (in dimensionless units). For times $t > 2\pi$ the wave number about the point $x = 0$ gradually decreases, so the amplitude of the disturbance at the center gradually increases (see Figures 4 b-d and 11 a-c). For larger times ($t >> 2\pi$) the size of the "central disturbance" increases very slowly since the group velocity $V$ of the wave packet becomes smaller and smaller as $k$ decreases in time ($V \approx k$). This may account for a major difference between tremor and earthquakes. The latter have a clear impulsive phase, reflecting the propagation of rupture with velocity approaching seismic velocities. In the case of tremor generated by the radiation mode of a frictional surface, the initial phase should gradually increase in time, and this is actually observed. The typical time for tremor to develop can be estimated from the relation $T \approx \pi/k^2$ (using formula (7) with $k^2 << 1$) or, in dimensional units, $T = \left(\frac{\pi}{k^2}\right) \cdot \left(\frac{b}{cA}\right) \approx 5$ s if $k^2$=0.1. So tremor can develop in this model over a time span of up to a few seconds.

Based on an analysis of tremor duration-amplitude scaling, Watanabe et al. (2007) concluded that the size of the source of tremor should be "scale-bound rather than scale-invariant". This result is consistent with the prediction of our model. Indeed, whatever the source exciting a local shear failure, the latter disturbance will quickly grow in size, but the "central disturbance" will grow very slowly (see formula (7)). Since tremor is generated within and in close proximity to this central disturbance, the size of the source should be scale-bound for given values of the parameters *b* and *A*.

The velocity and displacement spectra are the most important observational characteristics of tremor. Tremor triggered by earthquakes and ETS tremor have slightly different velocity spectra (see Figure 5b from the article Rubinstein et al, (2010)), i.e. the fall-off law is steeper for the former for the low frequency part of the curve. Visual comparison of the calculated spectra (dotted lines in Figures 8 and 14) with spectra observed for tremor triggered by earthquakes indicates that the closest match is for the spectra calculated with friction coefficients in the range 0.01 to 0.025. This observation allows us to estimate the typical lifetime *T* and the maximal size *L* of the emitted area of an individual signal initiated by ether a fast or slow source. For this range of friction coefficients, both lifetime and maximal size are in a the range from $5\pi$ to $10\pi$ (in dimensionless units), hence in a range from $T = 10\pi b/(cA) \approx 5$s to $T = 10\pi b/(cA) \approx 10$s and from $L = 10\pi b/A \approx 25$km to $20\pi b/A \approx 50$km. The size of the emitted area may be smaller if the distance from the initiation point to the boundary is smaller than the typical attenuation distance of a signal inside the fault. If the size of the emitted area is less than $2\pi$ the central frequency shifts to a higher frequency (Figure 10).

There is no essential difference between tremor produced by fast and slow initial failure (cf. Figures 8 and 14). Indeed, the central (smallest) frequency is basically the same in both cases. However the frequency content is visibly different for the highest frequencies. For the same values of friction coefficients the fast sources produce tremor with larger amplitude at higher frequency than the slow sources.

Zhang et al. (2011) analyzed the spectral content of Cascadia tremor between 2.5 and 20 Hz during an ETS event. They found that the displacement spectra have corner frequencies of about 3–8 Hz and an $f^{-2}$ to $f^{-3}$ at fall-off at higher frequencies. Figure 15 depicts the calculated displacement spectra produced by the fast and slow disturbances. The fall-off law for the higher frequencies is about $f^{-2.5}$ for both types of calculated tremor (see dotted lines at Figure 15, right panels), which is consistent with the observation. However only tremor from the fast disturbances exhibit a weak sign of a corner frequency, at ~6 Hz.

It has been observed that (1) a considerable portion of tremor signals include superposed LFE waveforms (Shelly et al., 2007(a); Brown et al., 2009) and (2) signals from VLF earthquakes are usually buried in tremor signals (Ito et al., 2007; 2009). What is the relationship between tremor and LFE and between tremor and VLF earthquakes? Our model implies that any failure inside a fault excites a radiation mode. Under appropriate conditions this radiation may propagate along the fault and produce tremor by the mechanisms described above. In this respect all types of earthquakes, including LFE and VLF earthquakes, can be sources of tremor. However, if the size of the failure area equals the size of the earthquake source, the size of the tremor source is much larger even though it originates in the same place. This may be a reason why the source of tremor is hard to determine.

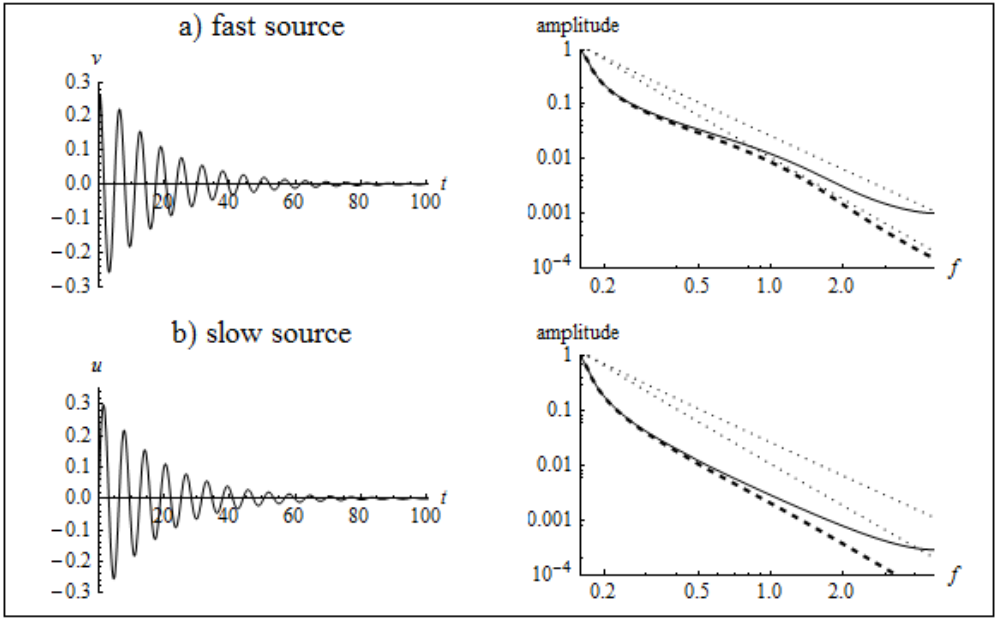

**Figure 15.** The calculated displacement (in arbitrary units), versus time (in dimensionless units). Also shown the corresponding displacement spectra in the range from $1/(2\pi)$ to $30/(2\pi)$. The reference fall-offs (dotted line) are $f^{-2}$ and $f^{-2.5}$. The dashed lines show the spectral density of the same signal but includes frequency-dependent attenuation during its propagation from the source to the measurement point. Computation made with $\alpha_s = \alpha_d = 0.01$.

There have been a few attempts to evaluate the effective seismic moment of tremor (Kao et al., 2010; Fletcher and McGarr, 2011). As already pointed out, tremor does not represent a rupture itself, as in an earthquake, but rather reflects an oscillatory disturbance on the frictional surface. Such oscillatory disturbances may exist under specific conditions intrinsic to the depth and fluid content of the fault, and may be excited by a rupture observed, for example, as a LFE. In this context the seismic moment, which depends on slip and slip area, may not much reflect the physics of the process.

Two mechanisms have been proposed to explain the appearance of tremor (e.g. Rubinstein et al., 2010): (1) tremor is generated by fluid-flow processes; (2) tremor reflects a frictional process (rupture) on a fault with low rupture speeds (Nakata et al, 2011). Both approaches require the presence of fluid and high fluid pressure. Our model does not directly require the presence of fluid in the source area, but for it to be consistent with observed tremor parameters, a comparatively small effective normal stress is required, which may not be expected at such depths without the presence of high pressure fluid. As our estimate showed, the resonant-type oscillations inside a fault may continue for up to a few dozen seconds. However it is known that the duration of a tremor burst may be as long as a few minutes, hours or even days. Thus, it is obvious that the source of tremor must function almost discontinuously. In this respect the approach developed here is complementary to existing models (e.g. Nakata et al., 2011; Ben-Zion, 2012), which suppose that tremor reflects a "frictional process with low rupture speed".

Essentially our model adds the hypothesis that resonant-type oscillations exist inside a fault. This addition may change our understanding of the nature of tremor in general, and the methods of its identification and location in particular.

Figures 8-10 and 14-15 demonstrate tremor initiated inside a fault by an idealized delta-function shape (fast) or Heaviside-function shape (slow) failure. Obviously, the shape of such idealized tremor is far from that observed data. Figures 16a and 16b shows the shape and the velocity spectra of the calculated tremor produced by 10 consecutive LFEs and VLF respectively, with slightly randomized amplitudes and time intervals between events. In these calculations we take into account frequency-dependent dissipation during propagation of the signal from the source to the surface. As one can see, tremor initiated by the fast sources include large amplitude spikes in contrast to tremor from the slow sources. The frequency content is also visibly different. The fall-off law for tremor from fast source changes from $f^{-1}$ to $f^{-2}$, whereas for tremor from slow sources it is about $f^{-1.5}$.

**Conclusion**

A Frenkel-Kontorova-type model has been developed for the quantitative description of ambient and triggered deep non-volcanic tremor. In a sense this model is complementary to the "shear faulting" type of model (e.g. Nakata et al., 2011; Ben-Zion, 2012). However, in our model tremor is not simply a collection of small slow earthquakes, but rather a result of resonant-type oscillations initiated by local failure such as a regular or slow earthquake or aseismic slip. The oscillations can exist only if the effective normal stress is very low. This is consistent with the observation that distant earthquakes and even tidal variations trigger tremor.

The specific features predicted by our model are: (1) the central frequency of tremor relates directly to the effective normal stress at the source location; 2) the upper limit of the size of the emitting area is a few dozen km; (3) tremor accompanies aseismic slip and earthquakes, including LFE, VLF and slow earthquakes; (4) the frequency content of tremor depends on the initial failure, with fast sources such as regular earthquakes and LFEs producing large amplitudes at higher frequencies in contrast to slow sources such as VLF earthquakes and aseismic slip; (5) there is no particular dependence of tremor characteristics on the frequency of the triggering seismic wave.

Our model also leads to the conclusion, in agreement with the conventional point of view, that observed tremor is a composition of shorter signals with duration of 5 to 10s (see Figure 16). The only difference between our approach and the conventional one is that we assume that the rupture of small asperities (which may be seen as LFE or VLF or may not be seen at all) initiates oscillations whose lifetime is longer than that of LFE. To explain the duration (minutes to hours) of tremor we also assume that observed tremor is a composition of multiple events. In this paper we basically consider one such event under idealized conditions. Although we do compare calculated and observed spectra, we do not intend or expect our model to describe quantitatively well all observed tremor parameters. The goal here is to show that such mechanisms are possible and that in the framework of our model there is a visible difference between tremor from fast and slow sources.

This model could be examined in the following ways. (1) The precise location of a tremor source is an ongoing problem (e.g. La Rocca, 2009; Rubinstein et al., 2010; Ide, 2012). The methods developed heretofore are based on the supposition that the source is a small area (essentially a point). Our model predicts that the actual emitting area may span dozens of km. If so, the procedure of source location should be essentially modified. (2) Since the central frequency of tremor depends on the effective normal stress, its value should vary with the location of the source. So there may be a slight difference in frequency between tremor from different depths of the same fault. This feature could be examined straightforwardly. (3) A comparison of the spectral composition (especially in the high frequency range) of tremor associated with a clearly identified slow earthquake and a LFE could be quite informative.

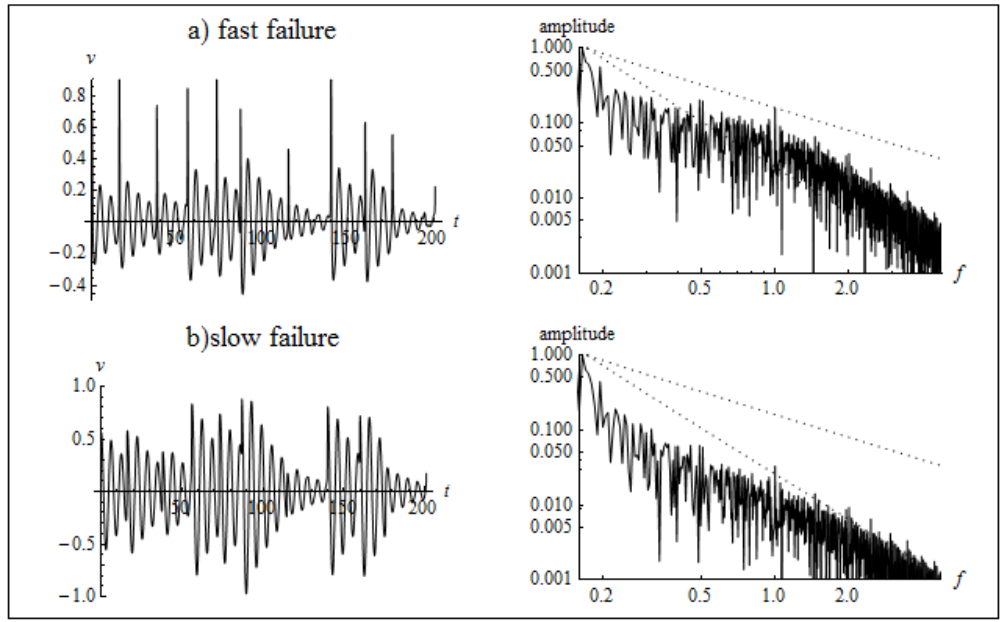

**Figure 16.** Calculated tremor (right panel) produced by 10 consecutive LFEs (a) and VLFs (b) with randomized amplitudes and time intervals between events. The reference fall-off is $f^{-1}$ and $f^{-2}$, respectively (dotted lines). Computation made with $\alpha_s = \alpha_d = 0.01$. The signal attenuation due propagation from source to surface have been included for both the v(t) and v(f) curves. To do this we multiplied the velocity spectrum by $\exp(-\pi f t/Q)$, where the travel time $t = 15$ s and $Q = 350$. We then recalculated the v(t) curves using the inverse Fourier transform.

**Acknowledgement**

We thank J.-P. Ampuero, N. Beeler, Y. Ben-Zion, J.H. Dieterich, D. Shelly and T. Suzuki for useful discussions and comments. This work was supported by NSF grant EAR-1113578.

**Appendix A**

The unperturbed sine-Gordon (SG) equation is fully integrable, i.e. solutions may be expressed analytically in terms of known functions. However the solutions of equation (3) with a source term generally can be obtained only computationally. There are methods to find analytical solutions in the case of a small perturbation. We will use the results of perturbation analysis developed by McLauglin and Scott (1978) (hereafter referred to as MS). Here we are interested in phonon-like solutions (the zero-soliton case in term of MS). The solution $\vec{W} \equiv \begin{Bmatrix} u(x,t) \\ u_t(x,t) \end{Bmatrix}$ may be expressed through the Green function $G_c$ as (see section VI.1 from MS):

$$\vec{W} = \int_0^t \int_{-\infty}^{\infty} G_c(x,t|x',t')\vec{F}(x',t')dx'dt', \tag{A1}$$

where $G_c(x,t \mid x',t') = \frac{1}{2\pi}\int_{-\infty}^{\infty} d\phi \begin{pmatrix} \cos[\varphi(t-t')] & \frac{1}{\varphi}\sin[\varphi(t-t')] \\ -\varphi\sin[\varphi(t-t')] & \cos[\varphi(t-t')] \end{pmatrix} \exp[-i\phi(x-x')]$, where $\varphi^2 = \phi^2 + 1$ and $\vec{F}(x,t) = \begin{Bmatrix} 0 \\ \Sigma_S^0 - f \end{Bmatrix}$

Supposing that $\Sigma_S^0 - f = a_{init}\delta(x)\delta(t)$ and consider only the first term of the vector $\vec{W}$ we find that expression (A1), after trivial integration, reduces to the one-dimensional integral,

$$u(x,t) = \frac{a_{init}}{2\pi}\int_{-\infty}^{\infty} \frac{\sin(\varphi t)\exp(-i\phi x)}{\varphi} d\phi. \tag{A2}$$

Using the same procedure we obtain a similar expression for the slow initial failure case

$$\Sigma_S^0 - f = a_{init}\delta(x)\eta(t),$$
$$u(x,t) = \frac{a_{init}}{2\pi}\int_{-\infty}^{\infty} (\frac{1-\cos(\varphi t)}{\varphi^2})\exp(-i\phi x)d\phi. \tag{A3}$$